\documentclass[12pt] {article}
\usepackage{graphicx}
\newcommand{{\y}}{\'{\i}}
\nonstopmode
\begin{document}

\begin{center}
{\bf MAGNETIC BEHAVIOR OF A MIXED ISING FERRIMAGNETIC MODEL IN
AN OSCILLATING MAGNETIC FIELD.}\\
\vskip 10 truept
G.~M.~Buend{\y}a, E.~Machado\\
\vskip 10 truept
\noindent Departamento de F{\y}sica. Universidad Sim\'on Bol{\y}var,\\
Apartado 89000, Caracas 1080, Venezuela.
\end{center}

\vskip 15 truept
\begin{center}
\noindent {\bf ABSTRACT}
\end{center}

The magnetic behavior of a mixed Ising ferrimagnetic system on a
square lattice, in which the two interpenetrating square
sublattices have spins $\sigma$ ($\pm 1/2$) and spins $S$ ($\pm
1,0$), in the presence of an oscillating magnetic field has been
studied with Monte Carlo techniques. The model includes nearest
and next-nearest neighbor interactions, a crystal field and the
oscillating external field. By studying the hysteretic response of
this model to an oscillating field we found that it qualitatively
reproduces the increasing of the coercive field at the
compensation temperature observed in real ferrimagnets, a crucial
feature for magneto-optical applications. This behavior is
basically independent of the frequency of the field and the size
of the system. The magnetic response of the system is related to a
dynamical transition from a paramagnetic to a ferromagnetic phase
and to the different temperature dependence of the relaxation
times of both sublattices.

\vskip 15 truept
\begin{center}
\noindent {\bf INTRODUCTION}
\end{center}

The behavior of ferrimagnetic compounds in the presence of
oscillatory fields have long been used for technological
applications such as high-density magneto-optical recording
~\cite{exp1}, but little is known about the mechanisms responsible
for this behavior. In a ferrimagnet the different temperature
dependencies of the sublattice magnetizations raise the
possibility of the appearance of compensation temperatures:
temperatures below the critical point, where the total
magnetization is zero ~\cite{Neel,Chavan}. It has been shown
experimentally that the coercive field is very strong at the
compensation point favoring the creation of small, stable,
magnetic domains ~\cite{Hansen}. This temperature dependence of the coercivity near the compensation point can be applied to writing and erasing in high-density magneto-optical recording media, where the temperature changes  are achieved by local heating the films by a focused laser beam. It has been shown that magneto-optic thin films with compensation temperatures higher than room temperatures can attain a direct overwrite capability \cite{Shieh}. As far as we know there
has been only very few crude attempts to reproduce theoretically the increase of the coercivity near the compensation point
using mean-field approaches ~\cite{Mansuripur}.Recently new classes of magnets are being synthesized with molecular organic chemistry techniques ~\cite {Turnbull}. Biocompatible, organic materials, optically transparent, with spontaneous moments at room temperature are not far from reality. Ferrimagnetic ordering seems to play a fundamental role in some of these materials. Ferrimagnetic compounds called Prussian blue analogs, with a critical temperature of 240 K have been reported ~\cite {Mallah}. Organometallic compounds as the amorphous $V(TCNE)_xy(solvent)$ where TCNE is tetracyanoethylene are believed to have ferrimagnetic structure and ordering temperatures as high as 400 ~\cite{Manriques}. Some of these compounds have compensation temperatures near 30 K ~\cite{Mathoniere}. Most of these compounds have been synthesized by assembling molecular building blocks of different magnetic moments in such a way that adjacent magnetic moment are antiparallel ~\cite{Turnbull}. Since real ferrimagnets have extremelly complicated structures mixed Ising models
have been introduced as simple systems that can show ferrimagnetic
behavior ~\cite{mixed,mixed2,Ohkoshi1} and may show compensation
points when their Hamiltonian includes second-neighbor
interactions ~\cite{Buendia1}. In this article, we present a Monte
Carlo study of a mixed Ising spin system, where spins that can
take the values $\pm 1/2$, and spins that can take the values $\pm
1, 0$, are nearest neighbors on a 2-dimensional square lattice and
interact antiferromagnetically. Spins of the same type are
next-nearest neighbors. We analyze the magnetic response of this
system in the presence of an oscillating magnetic field. From
these studies, we determine the dynamic order parameter, the
coercive field and their variation with temperature, frequency and
amplitude of the applied field, and size of the system. The
results reproduce the rapid increase of the coercitivity at the
compensation temperature. The dynamical order parameter
calculations suggest that the model exhibits a phase transition
between a paramagnetic and a ferromagnetic region. A similar
result was observed by a mean-field study of a simpler version of
this model ~\cite{Machado}. Mean-field approaches and Monte Carlo
simulations indicates the presence of a dynamical phase transition
in a kinetic Ising model ~\cite{Tome, Lo}. However the distinctive
behavior of the coercive field at the compensation temperature in
ferrimagnets seems to be related to the different relaxation times
of the sublattices. \vskip 15 truept
\begin{center}
\noindent {\bf THE MIXED ISING MODEL}
\end{center}

Our model consists of two interpenetrating square sublattices. One sublattice
has spins $\sigma$ that can take two values $\pm 1/2$, the other sublattice
has spins $S$ that can take three values, $\pm 1,0$. Each $S$ spin has only
$\sigma$ spins as nearest neighbors and vice versa.

The Hamiltonian of the model is given by,
\begin{equation}
{\cal H}=-J_1\sum_{\langle nn\rangle} \sigma_i S_j -
J_2\sum_{\langle nnn\rangle} \sigma_i \sigma_k +
D\sum_j S_j^2 - H(t)\left(\sum_i \sigma_i + \sum_j S_j\right)
\label{hamilt}
\end{equation}
where the $J$'s are exchange interaction parameters, $D$ is the crystal
field and $H$ is an oscillating magnetic field of the form,
\begin{equation}
H(t)=H_{0} cos(\omega t)
\end{equation}
where $\omega$ is the frequency of the external field, its period is given by,
$\Theta=2\pi/\omega$.  The $J$'s, $D$ and $H_0$ are all in energy units.
We choose $J_1$$=$$-1$ such that the coupling between nearest neighbors is
antiferromagnetic.

Previous results with Monte Carlo and Transfer-Matrix techniques have shown
that the $J_1$$-$$D$ model ($J_2$ and $H$ are equal to zero) does not have a
compensation temperature. These studies show that a compensation temperature
is induced by the presence of the next-nearest-neighbor (nnn) ferromagnetic
interaction, $J_2$, between the $\pm 1/2$ spins. The minimum strength of the
$J_2>0$ interaction for a compensation point to appear depends on the other
parameters of the Hamiltonian ~\cite{Buendia1}.

\vskip 15 truept
\begin{center}
\noindent {\bf MONTE CARLO CALCULATIONS}
\end{center}

We use standard importance sampling techniques to simulate the model described
by Eq.~(\ref{hamilt}) on a $L\times L$ square lattice with periodic boundary
conditions. Configurations are generated by randomly choosing spins on the
lattice and flipping them one at a time according to a heat bath algorithm.
In each complete sweep through the lattice $L\times L$ sites are visited.
Each Monte Carlo step per spin is associated to a time interval, $\tau_{S}$,
such that the frequency of the external field can be written as,
\begin{equation}
\omega = \frac{2\pi}{(NMCS)\tau_{S}}
\end{equation}
where $NMCS$ is the number of Monte Carlo steps per spin necessary to cover
an entire cycle of the field. To perform the simulations we arbitrarily
choose $\tau_{S}$ to be one, such that $\Theta=NMCS$. Our program calculates
the sublattice magnetizations per site at the time $t$ defined as,
\begin{equation}
M_1(t)=\frac{2}{L^2}\sum_j S_j(t)\ \ ,\ \
M_2(t)=\frac{2}{L^2}\sum_i \sigma_i(t)
\end{equation}
and the total magnetization per spin at the time $t$,
$M(t)=\frac{1}{2}\left[M_1(t)+M_2(t)\right]$.
The averages are taken over all configurations, the sums over $j$ are over
all sites with $S$ spins, and the sums over $i$ are over all sites with
$\sigma$ spins. Each sum has $L^2/2$ terms.

The compensation temperature is defined as the temperature below the critical,
$T_{\rm comp}<T_{\rm crit}$, where the two sublattice magnetizations cancel
each other such that the total magnetization is zero, i.e.,
\begin{equation}
|M_1(T_{\rm comp})|=|M_2(T_{\rm comp})|
\label{cond1}
\end{equation}
and
\begin{equation}
{\rm sign}[M_1(T_{\rm comp})]=-{\rm sign}[M_2(T_{\rm comp})].
\label{cond2}
\end{equation}

To characterize the time behavior we calculate the dynamical order parameter
$Q$ defined as,
\begin{equation}
Q=\frac{2\pi}{\omega} \oint {M(t)dt}.
\end{equation}
The closed integral implies that the integral is performed over a cycle of
the external magnetic field.

\vskip 15 truept
\begin{center}
\noindent {\bf RESULTS}
\end{center}
The value of Q is calculated by averaging its values over 100 cycles of the
external field, once the system is in its stationary state. Most of the
measurements were done for a $L=40$ lattice. Lattices of different sizes were
used to study the finite-size effects. In Fig. 1 we show a hysteresis loop,
$M(t)$ vs $H(t)$, for a particular combination of parameters in the
Hamiltonian. The coercive field $H_c$ is defined as the minimum value of the
external field needed for the total magnetization to go to zero, as is
indicated in the figure. In Fig. 2 and Fig. 3 we show the coercive field vs
the temperature for oscillating fields of several amplitudes, $H_0$. In the
same figures we also plot the total magnetization for the equivalent system
subject to a constant field of magnitude $H_0$. Notice that the compensation
temperature, defined as the point where the total magnetization is zero, previous verification that Eq. 5 and 6 are satisfied, increases with the magnetic field, whereas the temperature at
which the magnetization becomes discontinuous does the opposite. At a certain
field, which amplitude depends on the parameters of the Hamiltonian, both
temperatures become equal and for any field of larger amplitude there is no
more compensation point, as can be seen in Fig. 3(b). From the figures it is
clear that the coercive field increases in the vicinity of the compensation
temperature where it reaches its maximum . These results are summarized in
Fig. 4. As expected, the maximum value of the coercive field at the
compensation temperature is given by $H_0$.

It is interesting to notice the asymmetric behavior of the coercive field
around the compensation point. In the low temperature region, $T<T_{\rm comp}$,
the coercive field decreases with increasing $T$ until it reaches a minimum,
after which grows rapidly reaching its maximum at $T_{\rm comp}$, when
$T>T_{\rm comp}$ the coercive field decreases. Notice that for small values of
$H_0$ there is a range of temperatures for which the  coercive field is not
defined (see Fig. 4).
This behavior of the coercive field has been observed experimentally
~\cite{Ostorero}. This result can be understood by looking
at Fig. 5 were it is shown how the hysteresis loop changes with the
temperature. As the temperature increases the loop moves in such a way that
the coercive field increases until it reaches its maximum, after which, if the
temperature keeps increasing, the loop stays below (or above) the $M=0$ axis
without crossing it, meaning that the applied field is not strong enough to
flip the spins. If we look at Fig.6 where we plot the coercive field and the
dynamical order parameter vs the temperature we see that there is a dynamical
phase transition between a paramagnetic region, $Q \approx 0$ and a
ferromagnetic region $Q \ne 0$, the region where the coercive field is not
defined is well into the ferromagnetic phase where the magnetization does not
changes sign.

By changing the size of the system and studying the behavior of the coercive field, see Fig. 7, we notice that there are finite size effects, particularly evident for small systems ($L<20$). However, for larger systems, the location of the peak of the coercive field around the compensation temperature seems to be independent on the size of the system. For small systems ($L<20$) the peak of the coercive field appears before the system reaches its compensation temperature. Also the coercive field seems to decrease more rapidly for the larger systems.

In Fig. 8 we present some results that show the dependence of the
coercive field with the size of the system, these results agree
qualitatively with the experimental behavior of magnetic films and
nanostructured Fe and Ni samples ~\cite{nano} for which the
coercitivity depends on the average size of the grain. The size
dependence of the coercive field is very similar to the size
dependence of the switching field of a kinetic Ising model (field
at which magnetization reversal is thermally induced on
experimental timescales for given temperatures and system sizes),
which behavior has been shown to be strongly dependent on the
modes by which the system decays ~\cite{Richards, Per}.

Next, we explore how the results depend on the frequency. In Fig.
9 we show how the coercive field vs the temperature changes for
different values of the frequency of the external field. We found
a quite different response to the frequency of the magnetic field
depending on the dynamical phase of the system. In the
paramagnetic phase ($Q \approx 0$, see Fig. 6) the coercive field
is larger for systems driven by fields with higher frequency, but
in the ferromagnetic phase ($Q \neq 0$) just the opposite happens,
as can been seen in Fig. 9. This behavior is related to the
temperature dependence  of the relaxation time in the different
regions ~\cite{Per,Bertotti}. In the ferromagnetic phase we
must take into account the relaxation time of both sublattices the
$\sigma$ and the $S$, whereas in the paramagnetic phase only the
$\sigma$ one is relevant because the $S$ lattice follows the field
with almost no delay ~\cite{Machado}. We also notice in Fig.9 that
when the field has a high frequency the maximum value of the
coercive field (that occurs at the compensation temperature) does
not reach the field amplitude, i.e., the coercive field does not
reach its saturation value.

In Fig. 10 we show the behavior of the coercive field vs the
inverse frequency for different values of the temperature. If the
field has a long period the coercive field seems to reach a value
that is independent of the frequency and depends on the
temperature. Again we expect that this behavior is related to the
temperature dependence of the relaxation times in the
different phases.

\vskip 15 truept
\begin{center}
\noindent {\bf CONCLUSIONS}
\end{center}
We have applied a Monte Carlo algorithm to the study of the
magnetic response of a mixed Ising ferrimagnetic model to an
oscillating magnetic field. We found that this model gives very
good qualitatively agreement with the magnetic behavior of real
ferrimagnets. It shows a rapid increase of the coercive field at
the compensation temperature, a crucial feature that makes
ferrimagnetic compounds extremely useful for thermo-optical
applications. It also reproduces qualitatively the dependence of
the coercitivity with the size of the sample observed
experimentally. The results show the existence of a dynamical
phase transition in which the mean-period averaged magnetization,
$Q$, changes from $Q \approx 0$ to a $Q \ne 0$. 
Work in progress indicates that, as recent studies shows is also the case for the kinetic Ising model ~\cite{Per, Bertotti}, some aspects of the hysteretic response as its dependence on the frequency and amplitude of the oscillating field, depends on the metastable decay mode. Explaining the different behavior in the different regimes depending on the nucleation mechanism (i.e., single-droplet or multidroplet ) by which the system decays \~cite{Machado2}.

\vskip 15 truept
\begin{center}
\noindent{\bf ACKNOWLEDGMENTS}
\end{center}
We are indebted to Mark Novotny and Per An Rikvold for many
insightful comments during the course of this work. G.~M.~B also
acknowledges the kind hospitality of the Supercomputer
Computations Research Institute of Florida State University at
Tallahassee, Florida.

\newpage

\begin{center}
\noindent {\bf FIGURE CAPTIONS}
\end{center}


\noindent {\bf Figure 1.} Hysteresis loop ($J_2=6$, $D=-1.9$, $k_BT=0.5$,
$\omega =\pi/30$). The coercive field is indicated.\\


\noindent {\bf Figure 2.} Coercive field and magnetization vs temperature.
($J_2=6$, $D=-1.9$, $\omega =\pi/100$). a)$H_0=0.1$ b)$H_0=0.5$.
Notice the discontinuity in the coercive field.\\


\noindent {\bf Figure 3.} Coercive field and magnetization vs temperature.
($J_2=6$, $D=-1.9$, $\omega =\pi/100$).
a)$H_0=0.8$ b)$H_0=2.1$. For this choice of parameters there is not
compensation point for $H_0 > 1$.\\


\noindent {\bf Figure 4.} Coercive field  vs  temperature
($J_2=6$, $D=-1.9$, $\omega=\pi/100$). The maximum value of
$H_{\rm c}$ is given by $H_0$.\\


\noindent {\bf Figure 5.} Hysteresis loop ($J_2=6$, $D=-1.9$, $\omega=\pi/100$,
$H_0=0.5$). Notice that for high temperatures there is no coercive field
(see Fig. 6).\\


\noindent {\bf Figure 6.} Coercive field and dynamical order parameter vs
temperature ($\omega=\pi/30$, $H_0=0.5$, $J_2=6$, $D=-1.9$).\\


\noindent {\bf Figure 7.} Coercive field vs temperature for different lattice
sizes ($\omega=\pi/100$, $H_0=0.5$, $J_2=6$, $D=-1.9$).\\


\noindent {\bf Figure 8.} Coercive field vs lattice size
($\omega=\pi/100$, $H_0=0.5$, $J_2=6$, $D=-1.9$).
The lines are guides for the eye.\\


\noindent {\bf Figure 9.} Coercive field vs temperature
($H_0=0.5$, $J_2=6$, $D=-1.9$).\\


\noindent {\bf Figure 10.} Coercive field vs inverse frequency ($H_0=0.5$,
$J_2=6$, $D=-1.9$). The lines are guides for the eye.

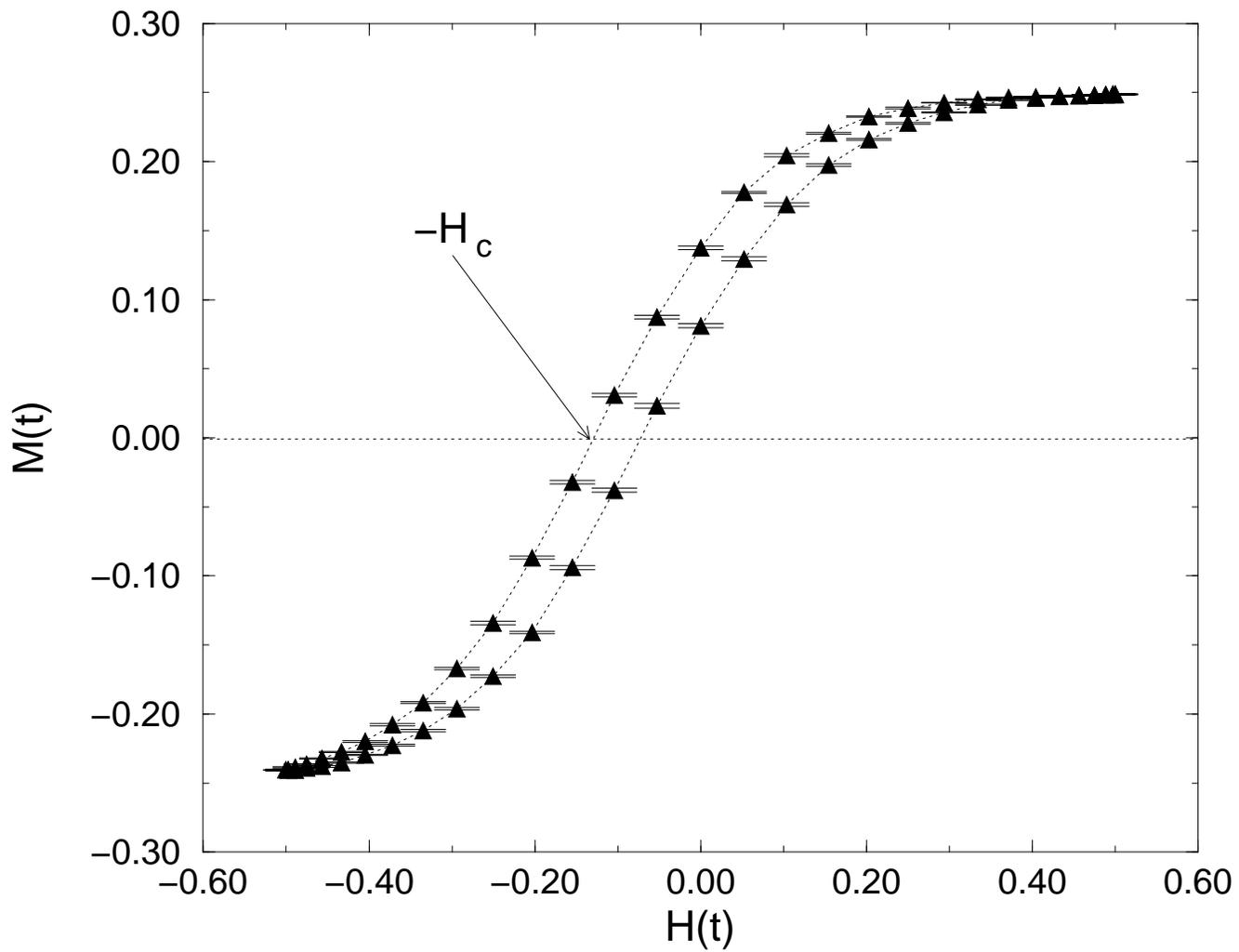
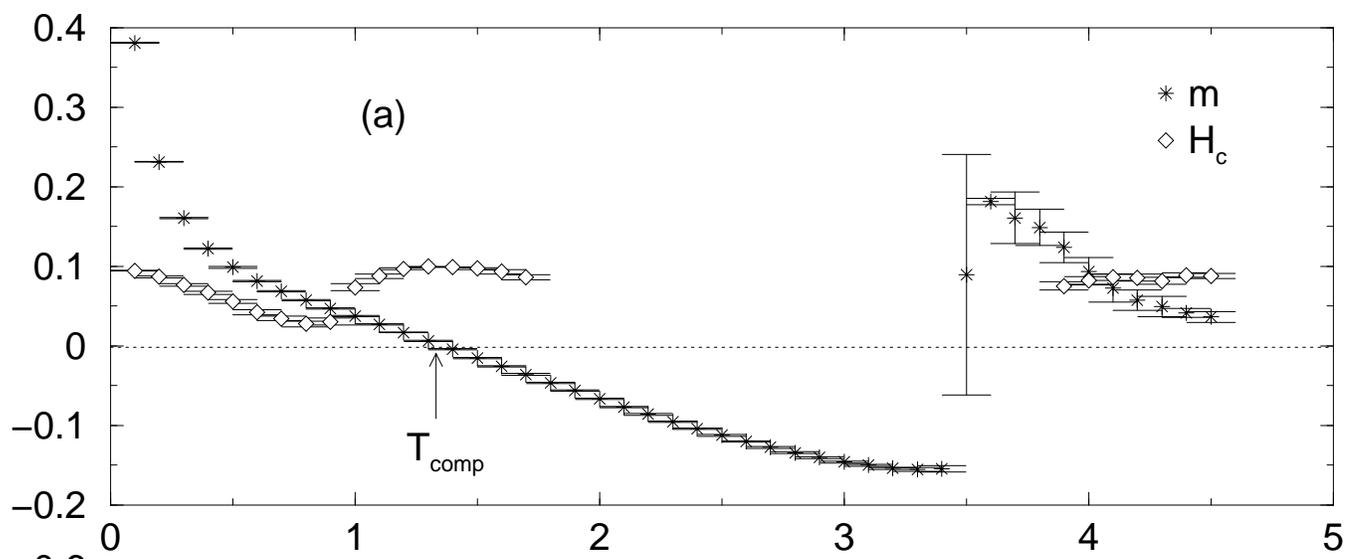
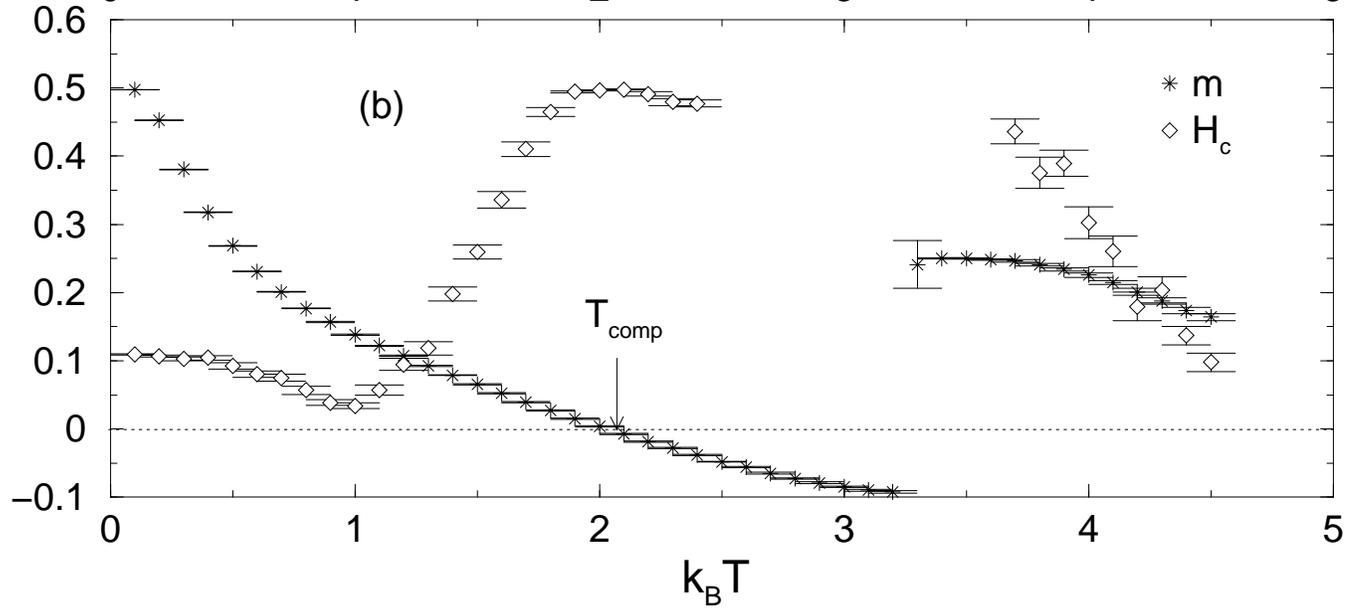
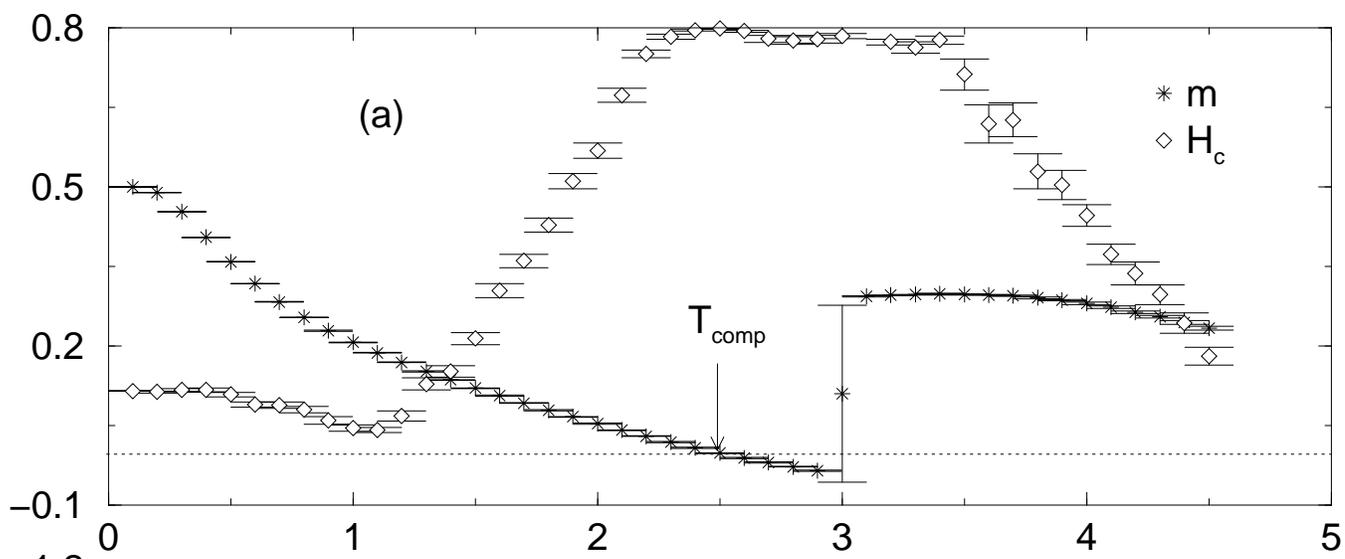
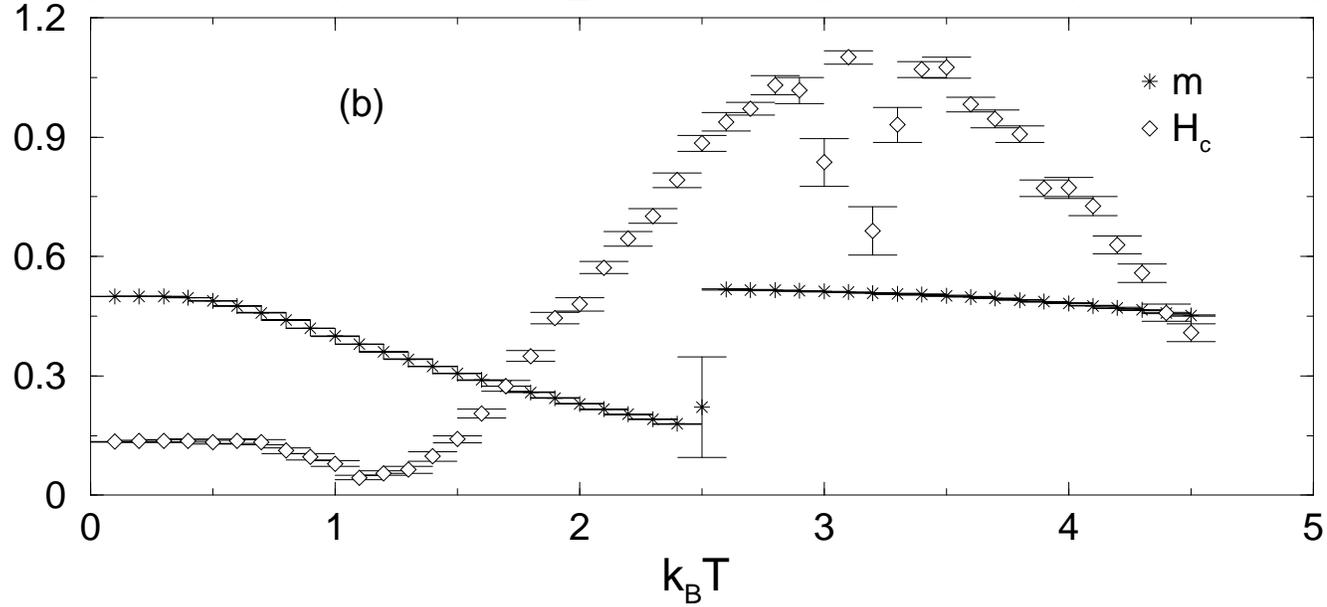
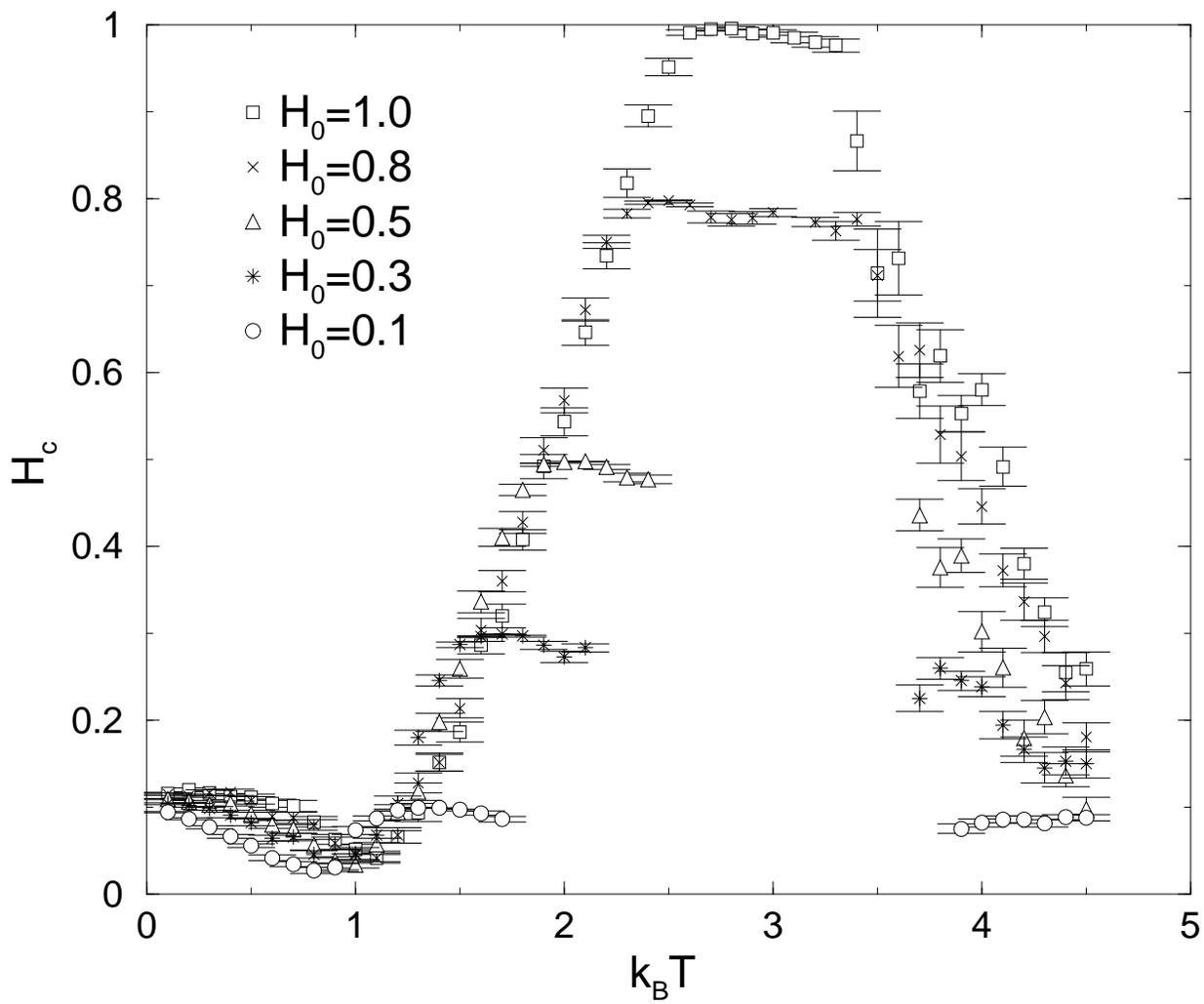
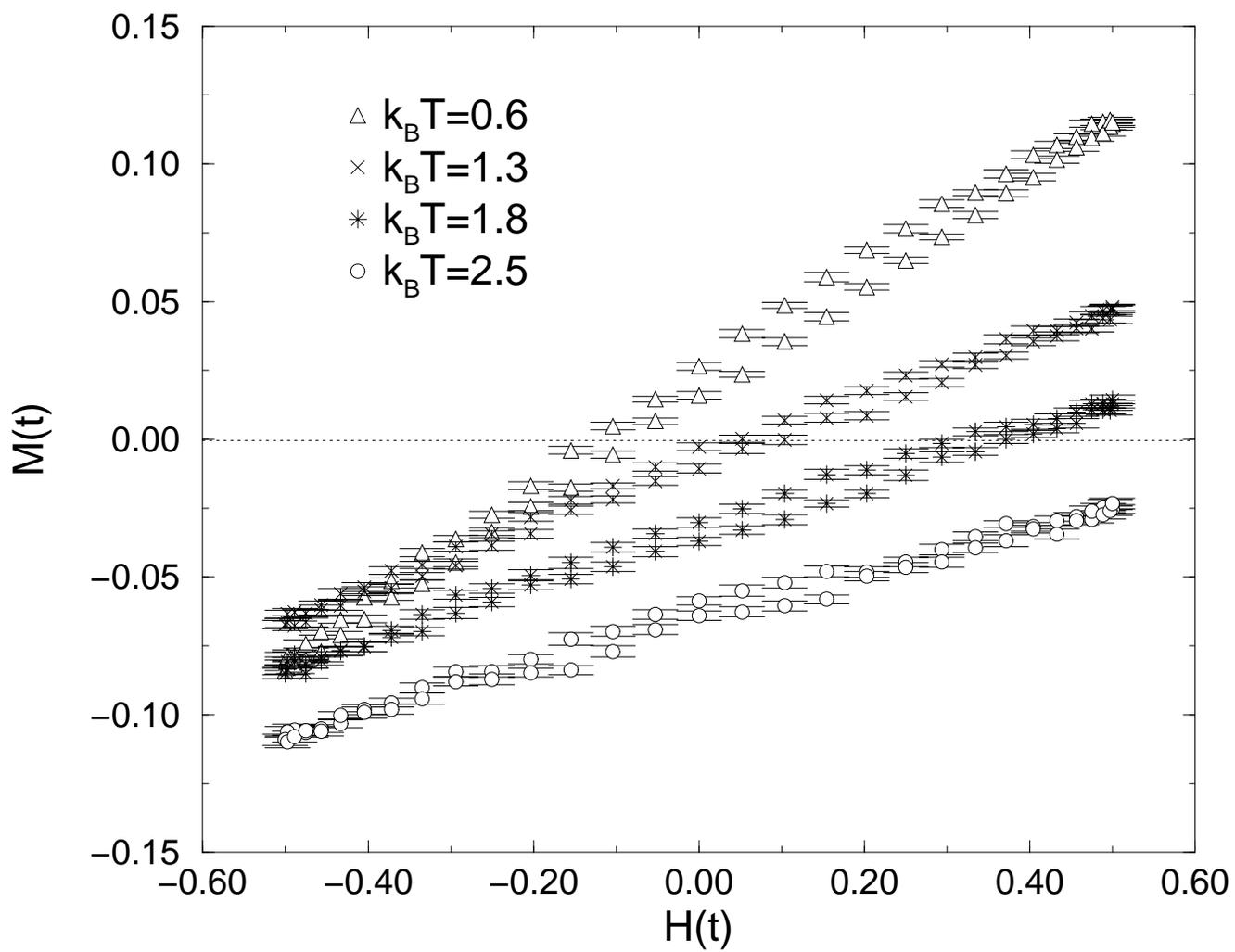
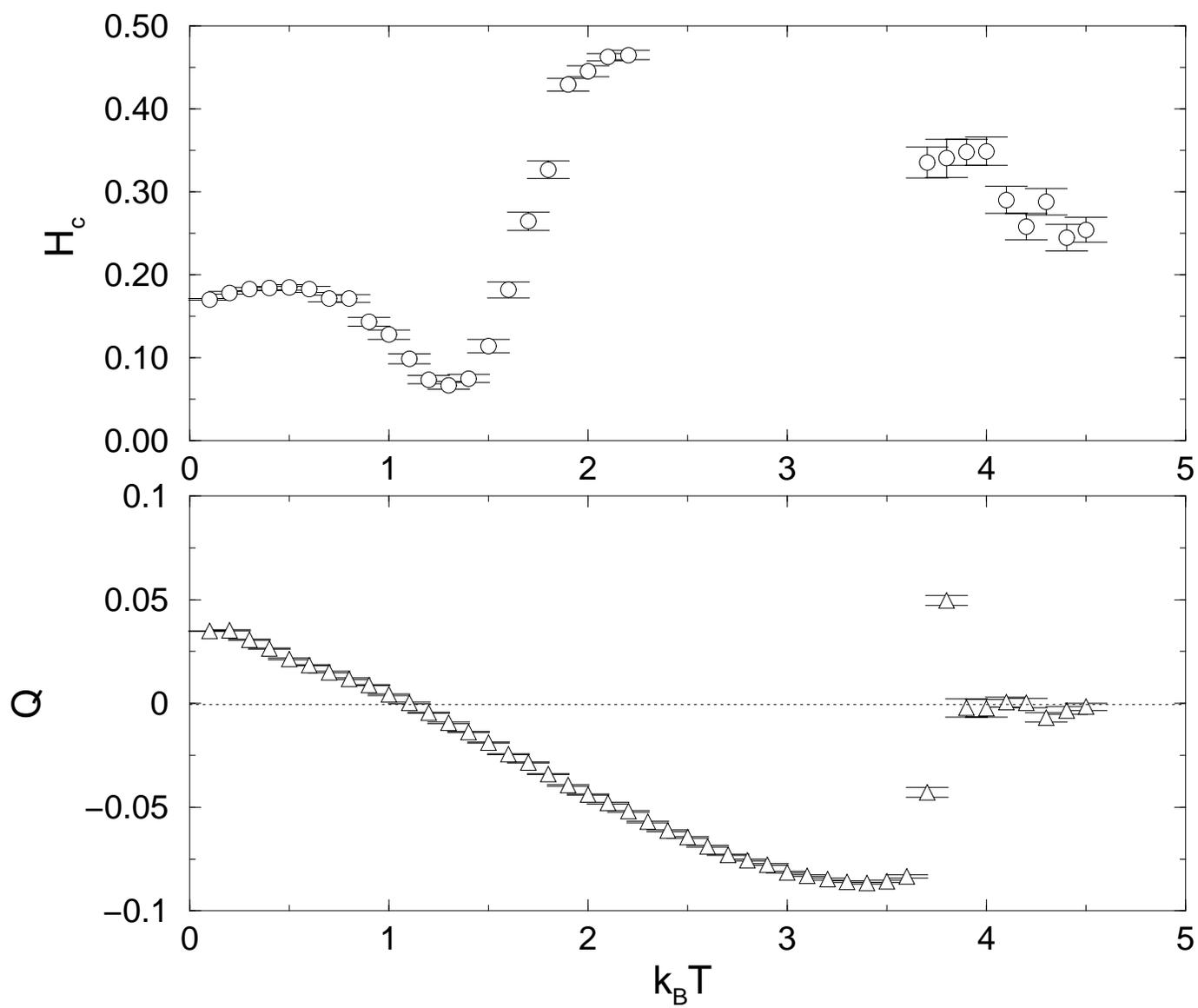
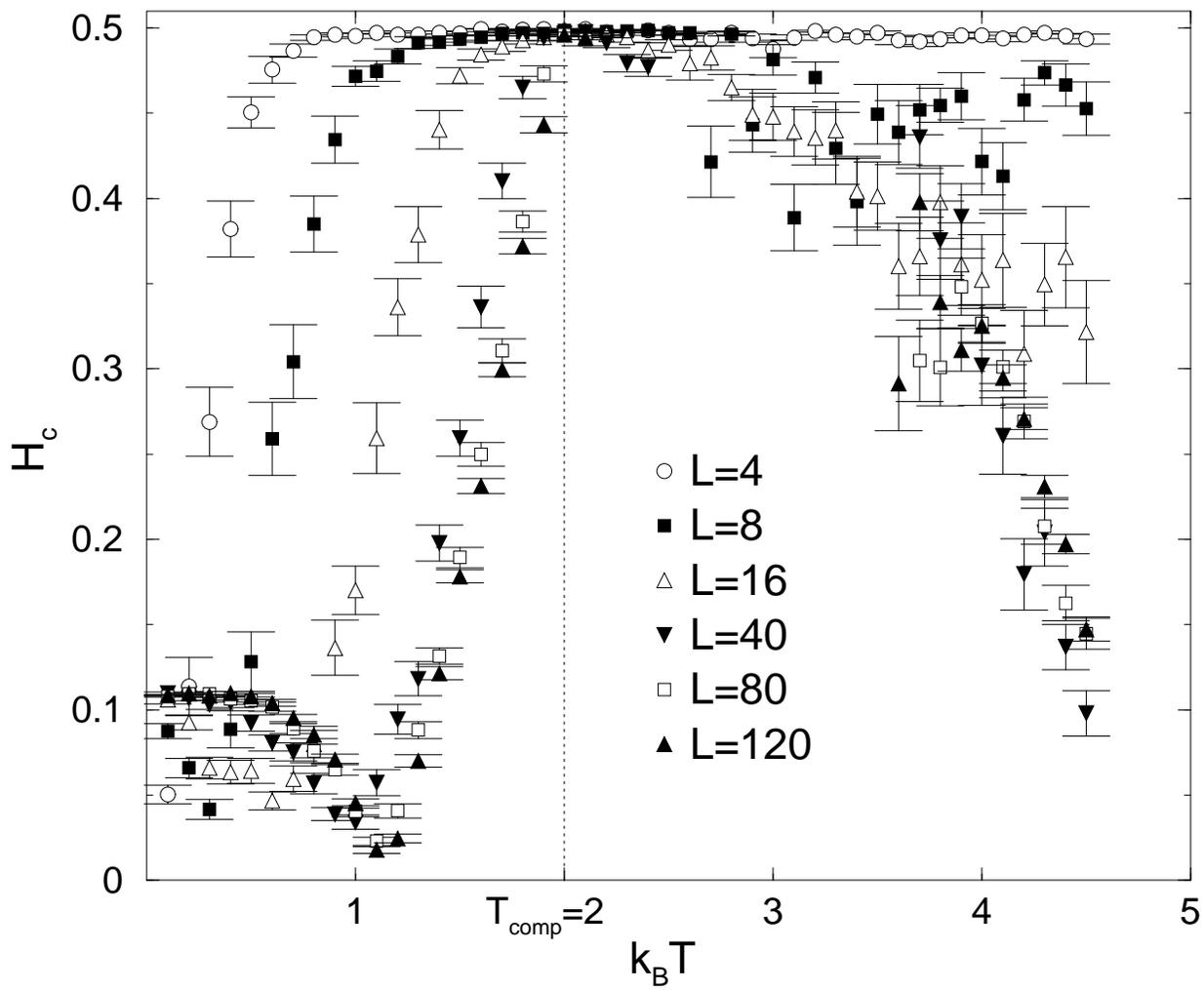
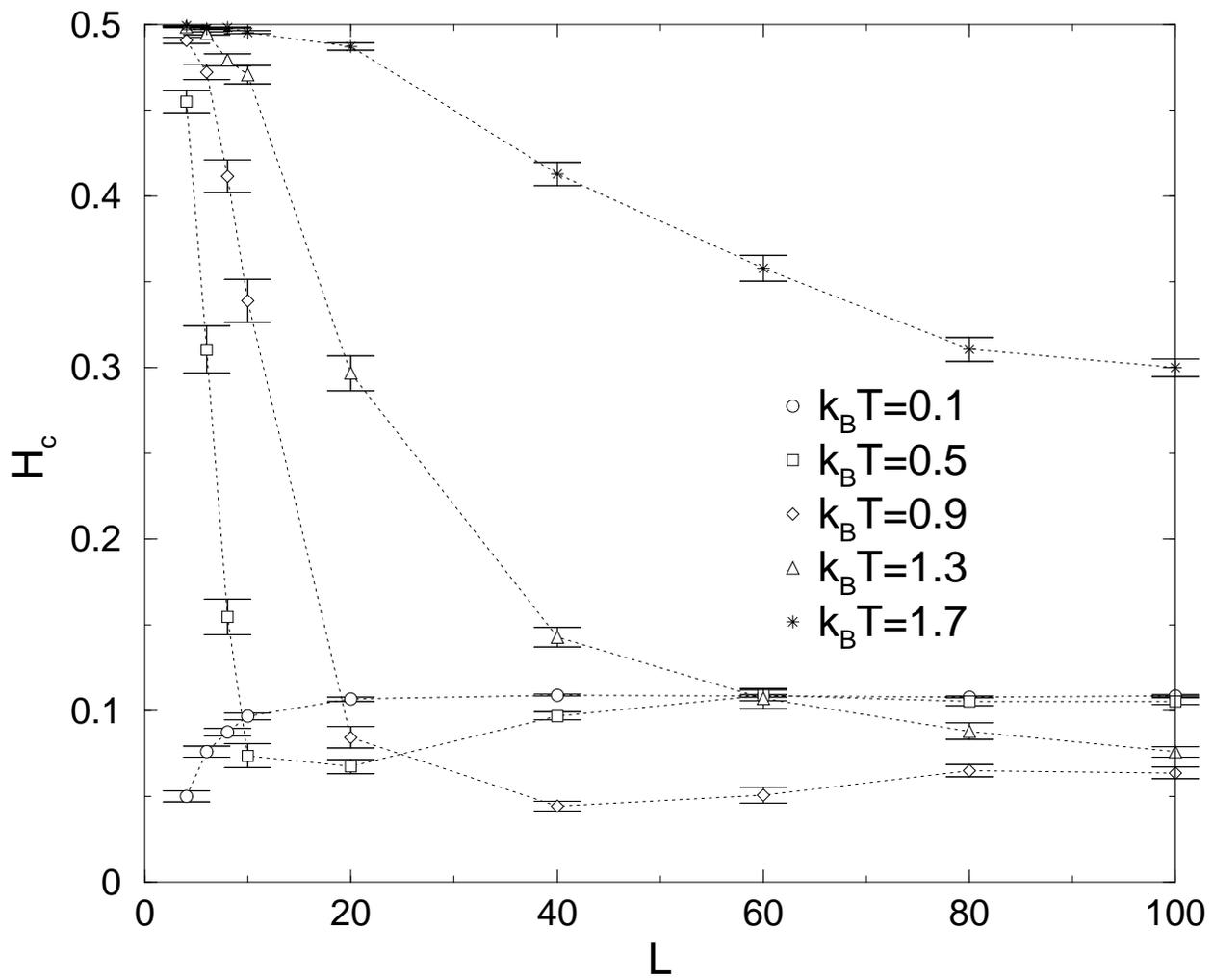
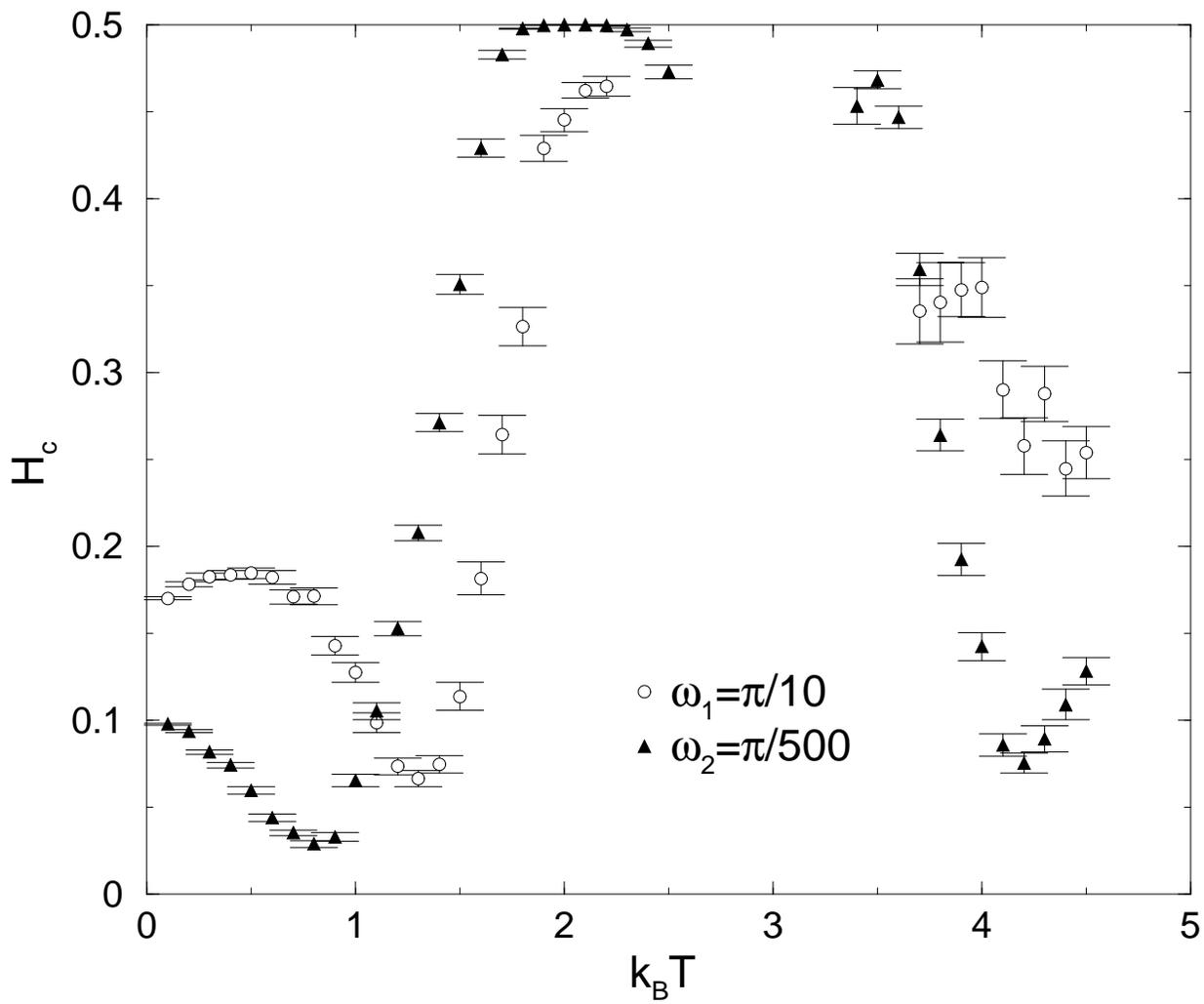
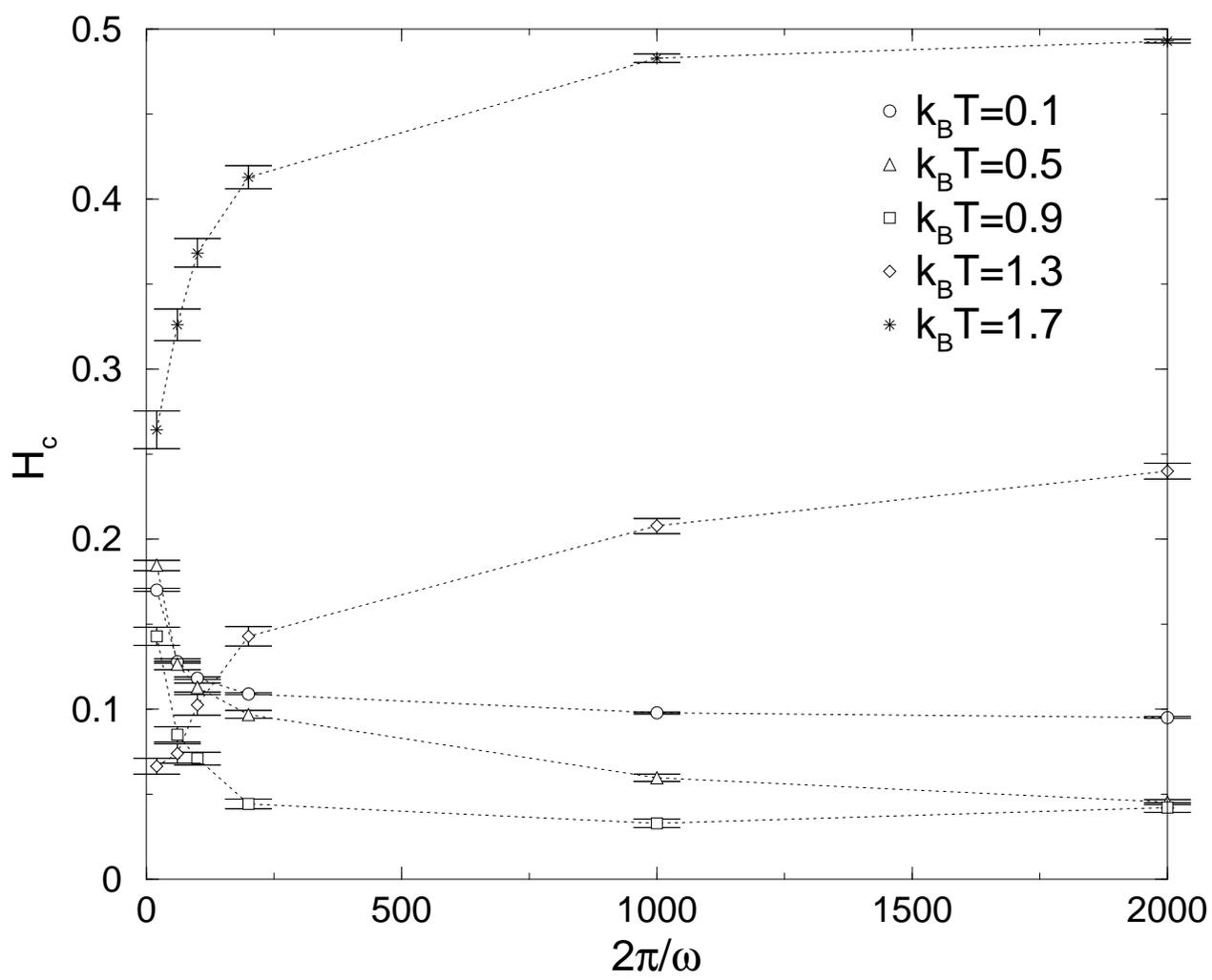
\end{document}